\documentclass[11pt]{article}
\usepackage{amssymb, amsmath, amsthm}
\usepackage{graphicx}
\newcommand{\beq}[1]{\begin{equation}\label{#1}}
\newcommand{\eeq}{\end{equation}}
\newcommand{\req}[1]{(\ref{#1})}
\newcommand{\bmu}[1]{\begin{multline}\label{#1}}
\newcommand{\emu}{\end{multline}}

\renewcommand{\[}{\left[}

\newcommand{\eq}{\triangleq}

\renewcommand{\S}{\mathcal{S}}
\renewcommand{\L}{\mathcal{L}}
\newcommand{\x}{{\textbf{\textit{x}}}}

\begin{document}
\begin{center}
{\Large\bf A simple construction of cover-free $(s,\ell)$-code with certain
constant weight}
\\[15pt]
{\bf A.G. D'yachkov, \quad I.V. Vorobyev, \quad N.A. Polyanskii,\quad V.Yu. Shchukin}
\\[15pt]
Moscow State University, Faculty of Mechanics and Mathematics,\\
Department of Probability Theory, Moscow, 119992, Russia,\\
{\sf agd-msu@yandex.ru,\quad vorobyev.i.v@yandex.ru,\quad nikitapolyansky@gmail.com,\quad vpike@mail.ru}
\end{center}

\medskip

{\bf Abstract.}\quad
In the given article we generalize a construction presented in \cite{mac96}. We give a method of constructing a cover-free $(s, \ell)$-code. For $k > s$, our construction yields a
$ {{n \choose s} \choose \ell}\times {n \choose k}$ cover-free $(s, \ell)$-code with a constant column weight.

\section{Notations, Definitions and Statement of Problems}
Let $N$, $t$, $s$ and $\ell$ be integers, where $1 \le s < t$ and $1 < \ell < t-s$.
Let $\eq$ denote the equality by definition, $|A|$ -- the size of the set $A$ and
$[N] \eq \{1, 2, \dots, N\}$ -- the set of integers from $1$ to~$N$. The standard symbol
$\lfloor a \rfloor$ will be used to denote the largest integer~$\le a$.

A binary $(N \times t)$-matrix
$$
X = \| x_i(j) \|, \quad x_i(j) = 0, 1, \quad i \in [N], \; j \in [t],
$$
with $N$ rows $\x_1, \dots, \x_N$ and $t$ columns $\x(1), \dots, \x(t)$ (\textit{codewords})
is called a \textit{binary code of length $N$ and size $t = \lfloor 2^{RN} \rfloor$},
where a fixed parameter $R > 0$ is called a \textit{rate} of the code~$X$.
The number of $1$'s in a binary column $\x = (x_1, \dots, x_N) \in \{0, 1\}^N$, i.e.,
\beq{w}
|\x| \eq \sum\limits_{i = 1}^N \, x_i, \qquad \x \in \{0, 1\}^N,
\eeq
is called a \textit{weight} of~$\x$. A code $X$ is called
a \textit{constant weight binary code of weight $w$}, $1 \le w < N$, if for any $j \in [t]$, the weight $|\x(j)| = w$.

\textbf{Definition 1.}~\cite{mp88, dv02}.\quad
A code $X$ is called a {\em cover-free $(s,\ell)$-code}
(briefly, {\em CF $(s,\ell)$-code}) if for any two non-intersecting sets
$\S,\,\L\subset[t]$, $|\S|=s$, $|\L|=\ell$, $\S\cap\L=\varnothing$,
there exists a row $\x_i$, $i\in [N]$, for which
\beq{property}
\begin{aligned}
&x_i(j)=0 \; \text{for any}\;  j\in\S,
\\
&x_i(k)=1\; \text{for any}\; k\in\L.
\end{aligned}
\eeq
\section{Main Result}
Let $\{{n \choose k}\}$ denote the family of $k$-subsets of set $[n]$, and $\{{{n \choose s} \choose \ell}\}$ denote the family of non-ordered $\ell$-tuples of distinct $s$-subsets of set $[n]$. For $s<k<n$ define a binary matrix $X(k,s,\ell, n)$ by letting the rows and columns be, respectively, represented by the members of $\{{{n \choose s} \choose \ell}\}$ and $\{{n \choose k}\}$ in the following fashion: for given $K\in \{{n \choose k}\}$ and given $\S=\{S_1,\ldots, S_{\ell}\}\in \{{{n \choose s} \choose \ell}\}$ the matrix $X(k,s,\ell, n)$ has a $1$ in its $(K,S)$-th entry if and only if there exists at least one $S_i\in \S$ such that $S_i\subset K$.

\textbf{Theorem 1.}\quad \textit{$X(k,s,\ell,n)$ is an ${{n \choose s} \choose \ell}\times {n \choose k}$ cover-free $(s, \ell)$-matrix with a constant column weight.}

\textbf{Proof.}\quad Consider arbitrary $\S$ and $\L\subset [{n \choose k}]$, $|S|=s$, $|\L| = \ell$, $\S\cap \L=\varnothing$. Let the corresponding elements of $\{{n \choose k}\}$ be, respectively, $K_1$, \ldots, $K_s$ and $K'_1$, \ldots, $K'_{\ell}$. Since all elements of  $\{{n \choose k}\}$ are
distinct and of the same size, we can find sets $S_1$,\ldots, $S_\ell$ such that $S_i\subset K'_i$ and $S_i\not\subset K_j$ for any $i\in[\ell]$ and $j\in[s]$. Therefore, the row which corresponds to the set $\{S_1,\ldots S_\ell\}$ satisfies cover-free $(s,\ell)$-condition~\req{property}. Observe that the matrix $X(k,s,\ell, n)$ is a constant-weight matrix because all columns of the matrix possess the same properties. \qed

One can check that we have to choose $k=n/2$ to maximize the size of code $X$. In asymptotic regime the parameters of $X(n/2, s,\ell, n)$ are:
$$
t = 2^{n(1+o(1)},\quad N=\frac{n^{s\ell}}{(s!)^{\ell}\ell!}(1+o(1)).
$$


\begin{thebibliography}{99}
\bibitem{mp88}
\textit{Mitchell C.J.},  \textit{Piper F.C.},
``Key storage in Secure Networks'',
\textit{Discrete Applied Mathematics}, v.~21, pp.~215-228,  1988.

\bibitem{dv02}
\textit{A. G. Dyachkov}, \textit{P. Vilenkin}, \textit{A. Macula}, and \textit{D. Torney}, ``Families of finite sets in which
no intersection of  sets is covered by the union of $s$ others'', \textit{J. Combin. Theory}. Ser. A,
99 (2002), pp. 195-218.

\bibitem{mac96}
A.\,J. Macula, A simple construction of $d$-disjunct matrices with certain constant weights. Discrete Math., Ser.~A,  \textbf{162}:1-3 (1996), 311–312.
\end{thebibliography}
\end{document}